\documentclass{amsart}

\usepackage[utf8]{inputenc}
\usepackage{amsmath, amssymb, amsbsy, amsthm, amsaddr}
\usepackage{graphicx, color, url}
\usepackage{scrtime, currfile}

\renewcommand{\phi}{\varphi}

\newcommand{\abs}[1]{\left|#1\right|}

\newcommand{\lpm}{\begin{pmatrix}}
\newcommand{\rpm}{\end{pmatrix}}

\begin{document}
\title{First attempts to model the dynamics of the Coronavirus outbreak 2020}
\author[Th.~Götz]{Thomas Götz}
\address{Mathematical Institute\\University of Koblenz--Landau\\D--56070 Koblenz}
\email[Th.Götz]{goetz@uni-koblenz.de}

\begin{abstract}
Since the end of 2019 an outbreak of a new strain of coronavirus, called 2019--nCoV, is reported from China and later other parts of the world. Since January 21, WHO reports daily  data on confirmed cases and deaths from both China and other countries~\cite{WHO20}. In this work we present some discrete and continuous models to discribe the disease dynamics in China and estimate the needed epidemiological parameters. Good agreement with the current dynamics has be found for both a discrete transmission model and a slightly modified SIR--model.

\end{abstract}

\maketitle

\section{Introduction}
\begin{figure}
\includegraphics[width=.7\textwidth]{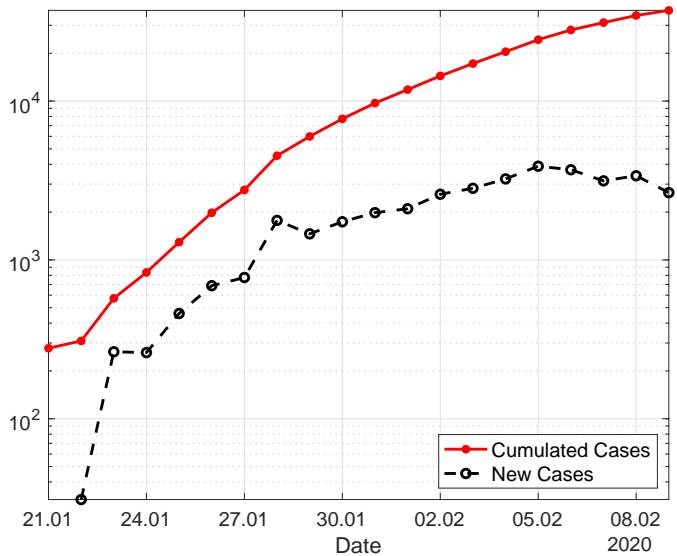}
\caption{\label{F:rawdata} Cumulated and new cases of Corona infections in mainland China.}
\end{figure}

In December 2019, first cases of a novel \emph{pneumonia of unknown cause} were reported from Wuhan, the seventh--largest city in China. In the meantime, these cases have been identified as infections with a novel strain of coronavirus, called 2019--nCoV. Its genome sequence turned out to be $75$-- to $80$--percent identical to the SARS--coronavirus, that caused a major outbreak in Asai in 2003.  At the beginning of January 2020, the virus spread over mainland China and reached other provinces. Increased travel activities due to the Chinese new year festivities supported the expansion of the infection. By mid of January, China reported a sharp rise in cases with about 150 new patients. From January 21 onwards, WHO's daily situation reports contain the latest figures on confirmed cases and deaths, see~\cite{WHO20}. Our work is based on these data for the mainland of China. The National Health Commission of the People's Republic of China also provides daily reports~\cite{NHCC}, however this website is only available in Chinese and hence we did not use it for our analysis. By $Y_k$ we denote the cumulated Corona cases in mainland China on day $k$, where $k=0$ corresponds to beginning of the data recordings on January, 21. In Figure~\ref{F:rawdata} we show the data until February, 9 in a semi--logarithmic plot. The new cases $N_k:=Y_k-Y_{k-1}$ are also given. The data for other countries than China is not included in our first modeling approach.

\section{Discrete Infection Models}

\begin{figure}
\includegraphics[width=.7\textwidth]{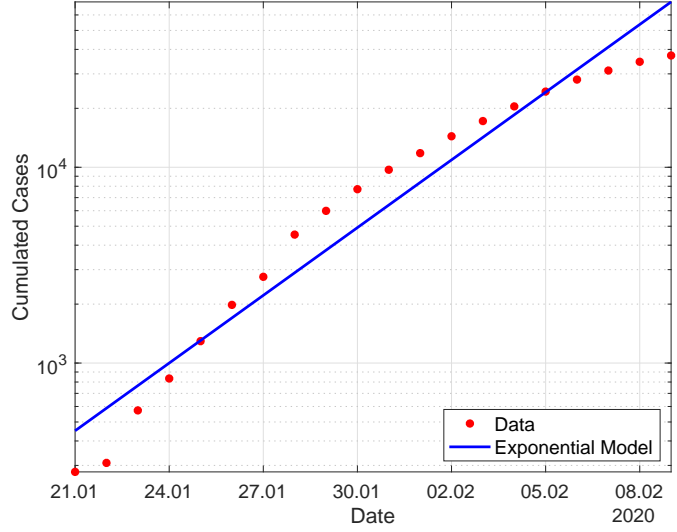}
\caption{\label{F:expmodel} Exponential model~\eqref{E:expmodel} compared to the data. The parameters are obtained from a logarithmic least squares fit.}
\end{figure}

Assuming, that the number of daily new infections is directly proportional to the total number of currently infected leads to a discrete exponential model
\begin{equation}
	\label{E:expmodel} \tag{Exp}
	y^\text{exp}_{k+1} = y^\text{exp}_k + r y^\text{exp}_k = (1+r)^{k+1} y_0
\end{equation}
To estimate the infection growth rate $r$ and the initial infected population $y_0$ at day $k=0$ (i.e.~Jan, 21), we use a least--squares fit to the observed data $(k, Y_k)$ for $k=0,\dots n$ with $n=15$ in a logarithmic version. We obtain the estimates
\begin{align*}
	\hat{r} &\simeq 0.304 \\
	\hat{y_0} &\simeq 451
\end{align*}
The coefficient of determination (in statistics usually called the $R^2$--value) of this fit is given by
\begin{equation*}
	R^2 = 0.9484
\end{equation*}
In Figure~\ref{F:expmodel} we visualize the exponential model~\eqref{E:expmodel} in comparison to the data. Although, the $R^2$--value is quite large, the observed behavior is far from being purely exponential. Especially in the last days, the disease dynamics has been slowing down. This is also reflected in the decreasing number of new cases, see Fig.~\ref{F:rawdata}. The exponential model is over--estimating the number of infected cases. 

\begin{figure}
\includegraphics[width=.7\textwidth]{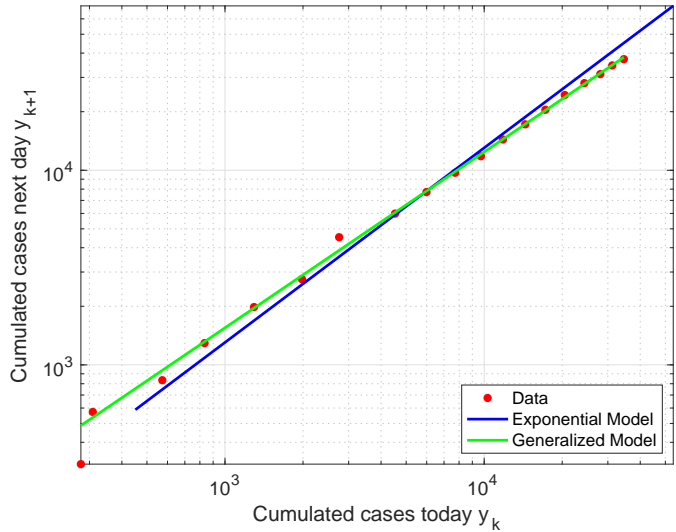}
\caption{\label{F:loglog1} Double--log plot of the cumulated cases tomorrow $y_{k+1}$ vs. the  cumulated cases today $y_k$.}
\end{figure}

This can be seen as well, when plotting the infected cases on the next day $Y_{k+1}$ vs.~the infected cases today $Y_k$ in a double logarithmic plot, see. Figure~\ref{F:loglog1}. In case of the exponential model~\eqref{E:expmodel}, we get
\begin{equation*}
	\ln y_{k+1} = \ln y_k + \ln (1+r)
\end{equation*}
a straight line of slope $1$ in the double log--plot. However, the real data reveals a slightly smaller increase. Therefore, we may generalize the model to
\begin{align}
	\label{E:genexp} \tag{Gen}
	y^\text{gen}_{k+1} = (1+c) (y^\text{gen}_k)^\beta
\intertext{or in the logarithmic version}
	\ln y^\text{gen}_{k+1} = \beta \ln y^\text{gen}_k + \ln (1+c) \notag
\end{align}
The parameters $\beta$ and $c$ can again be estimated from a least squares fit and we obtain
\begin{equation}
	\hat{\beta} = 0.904 \quad \text{and} \quad \hat{c}=2.02\;.
\end{equation}
In this fit, we have excluded the first data point $(\ln Y_1, \ln Y_2)$ as an outlier, cf. Fig.~\ref{F:loglog1}. A value of $\beta<1$ models an increase in the number of infected cases, that is below that standard exponential model~\eqref{E:expmodel} and that is in accordance with the reported data.

Another possible extension of the exponential model takes into account the effect of \emph{awareness} in the population. From a purely heuristic point of view, one can describe this by a nonlinear relation between the number of new infections $y_{k+1}-y_k$ and the current cases $y_k$, e.g.
\begin{equation}
\label{E:nonlinear} \tag{NonLin}
	y^\text{nl}_{k+1} - y^\text{nl}_k = \rho (y^\text{nl}_k)^\alpha 
\end{equation}
For $\alpha=1$, we recover the simple exponential model~\eqref{E:expmodel}. For exponents $\alpha<1$, the number of new--infections is reduced for high infection numbers to account for possible awareness in the population.

\begin{figure}
\includegraphics[width=.7\textwidth]{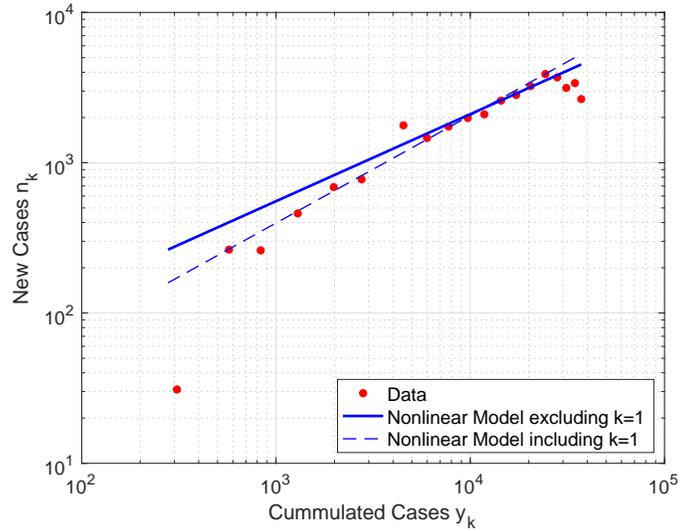}
\caption{\label{F:loglog2} Log--log--plot of the new cases $n_k$ vs.~cumulated cases $y_k$. The fits for the nonlinear model~\eqref{E:nonlinear} are excluding the first data point as an outliner (solid line) or include it (dashed line). }
\end{figure}

\begin{figure}
\includegraphics[width=.7\textwidth]{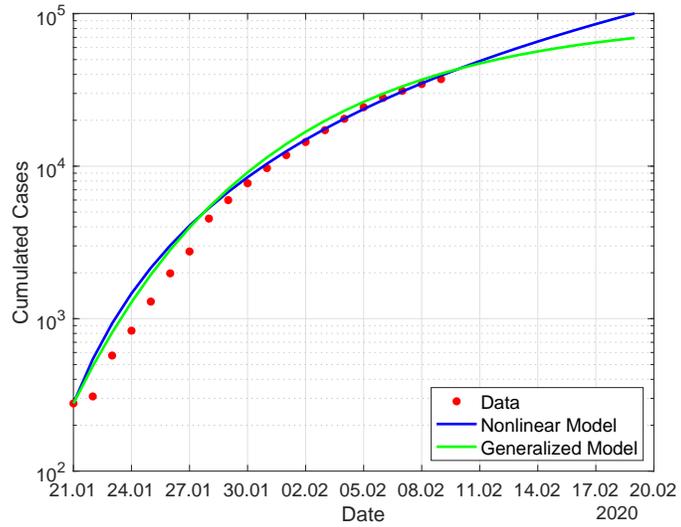}
\caption{\label{F:nonlin2}Data, nonlinear model~\eqref{E:nonlinear} (blue) and generalized exponential model~\eqref{E:genexp} (green).}
\end{figure}

For the given data (again \emph{excluding} the first data point as an outlier), we perform a least squares fit to the double logarithmic plot (see Fig.~\ref{F:nonlin2}) of the current cases $y_k$ vs.~the new infections $n_k$. This yields the parameter estimates
\begin{equation*}
	\hat{\rho} = 10.1321 \quad \text{and} \quad \hat{\alpha} = 0.5794
\end{equation*}
with an $R^2$--value of $0.9287$. Including the first data point (see dashed line in Figure~\ref{F:loglog2}) yields a fit with a smaller $R^2$--value of $0.8011$ showing a lower quality of the fit. Moreover, in this case, we would again drastically over--estimate the current behavior. Figure~\ref{F:loglog2} again shows the reduction of new infections in the last days. The last three data points significantly deviate from the previous behavior which can be described rather well by the nonlinear model. 

In Figure~\ref{F:nonlin2} we show the prediction based on the generalized model~\eqref{E:genexp}(green curve) and the nonlinear model~\eqref{E:nonlinear} (blue curve) compared to the observed data. The two models are extended 10 days beyond the current date to show their predictions.  On the time interval, where data is available, both models yield very similar estimates, but for future predictions, the generalized exponential model shows significantly lower case numbers. Based on the reduced number of new infection in the last days, it seems that the generalized model~\eqref{E:genexp} is better suited to describe the future disease dynamics. 

\section{SIR--Model}

In contrast to the above discrete models, we analyze in the sequel shortly a standard--SIR model. The total population equals to a reservoir of mainland China with $N=1.4\cdot 10^9$ inhabitants. The export of the disease to other countries is not taken into account. Due to the short time horizon of not more than $1$ month, effects of birth and natural death are excluded form the model. The time scale is measured in days, the recovery rates is assumed to be $\sigma=14^{-1}$ implying a recovery period of $14$ days. 

Let $I$ and $R$ denote the currently infected and recovered individuals. The susceptible individuals are given by $S=N-I-R$. The standard SIR--model without demographic terms leads to the two coupled ODEs
\begin{subequations}
\label{E:SIR1}
\begin{align}
	\frac{d}{dt} I &= \theta (N-I-R)\cdot I - (\mu+\sigma) I \\
	\frac{d}{dt} R &= \sigma I
\end{align}
\end{subequations}
where $\mu$ denoted the disease induced mortality rate. In addition, we introduce the cumulated number of infected $C_k=\int_0^{t_k} I(t)\, dt$.

\begin{figure}
\includegraphics[width=.7\textwidth]{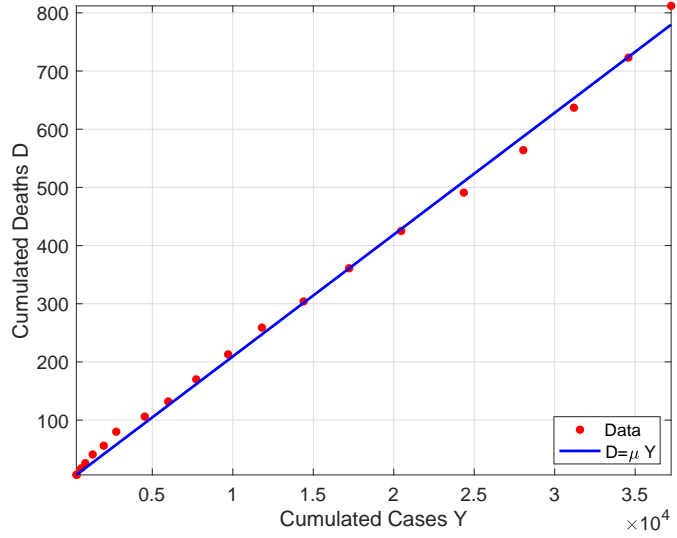}
\caption{\label{F:death} Observed cumulated deaths vs.~cumulated infections. The blue line shows the linear relation using a constant disease induced death rate $\mu$.}
\end{figure}

The cumulated number of deaths equals
\begin{equation*}
	D_k = \int_0^{t_k}\mu I(t)\, dt = \mu C_k
\end{equation*}
Using the data provided by WHO, we estimate the mortality rate 
\begin{equation*}
	\hat{\mu} = \frac{1}{n+1} \sum_{k=0}^n \frac{Z_k}{Y_k}
\end{equation*}
where $Z_k, Y_k$ denotes the cumulated observed death or infected cases at day $k$, see~Figure~\ref{F:death}. Based on the given date we obtain an estimated mortality rate $\hat{\mu}=2.09\%$. As can be seen from the graph, the assumption of a constant disease induced death rate shows rather good agreement with the given data. 

Using the above estimated death rate of approx.~$2\%$, we solve the SIR--model~\eqref{E:SIR} and compare its results for the cumulated infection cases $C$ to the reported data. In a first simulation, we use a constant force of infection $\theta$. Using a least--squares fit, we can estimate its value $\hat{\theta}$ form the observed data. However, during the corse of the disease, quarantine measures have been take to reduce the spread of the disease. In the SIR--model, such quarantine measures lead to a reduction in the force of infection. Therefore, we have also simulated a second model, where the force of infection is assume to be piecewise--constant
\begin{equation*}
	\theta(t) = \begin{cases}
		\theta_1 & \text{for } t<t_s \\
		\theta_2 & \text{for } t\ge t_s
		\end{cases}
\end{equation*}
where $t_s$ denotes the switching time. Again, we have performed a least--squares fit to estimate $\theta_1, \theta_2$ and $t_s$ based on the reported data. The fit is based on minimizing the $L_2$--difference
\begin{equation*}
	\text{$L_2$--diff} = \sum_{k=0}^n \abs{Y_k - C_k}^2
\end{equation*}
between the reported data $y_k$ and the simulated result $C_k$.

\begin{figure}
\includegraphics[width=.7\textwidth]{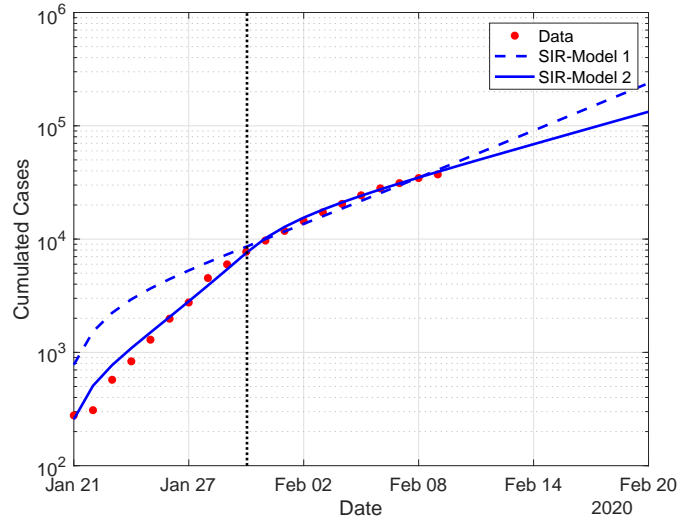}
\caption{\label{F:SIR} Simulations based on the SIR--model compared to the observed cumulated cases. The SIR--model 1 (dashed curve) uses a constant force of infection $\theta$. The SIR--model 2 (solid curve) incorporates a different forces of infection before the onset of quarantine effects (left of the vertical dotted line) and after (right of the vertical dotted line)}
\end{figure}

Figure~\ref{F:SIR} compares the simulation results of the two SIR--models to the given data. In Table~\ref{T:SIR} we summarize the results of the two SIR--models along with their respective basic reproductive numbers
\begin{equation*}
	R_0 = \frac{\theta}{\mu+\sigma}
\end{equation*}
For our modified SIR--model 2 with two different forces of infection $\theta$, the basic reproductive number based on the later value for $\theta$ equals to
\begin{equation*}
	\hat{R_0} = \frac{\theta_2}{\mu+\sigma} = 3.31\;.
\end{equation*}
Other sources, see~\cite{LW20, RBCea20} report a range between $2.1$ to $3.1$.

\begin{table}
\caption{\label{T:SIR} Parameters for the SIR--simulations as shown in Figure~\ref{F:SIR}.} 
\begin{tabular}{rlll}
& Parameter 	& SIR--Model 1 & SIR--Model 2 \\ \hline
Initial value & $I(0)$		& $778$		& $256$ \\[.2ex]
Force of infection & $\theta$	& $0.58$	& $\begin{cases} 2.12 & t<9 \\ 0.31 & t\ge 9 \end{cases}$\\[.2ex]
Switching time & $t_s$		& ---	 	& 30.01.2020 \\[.2ex]
Basic repro. number & $R_0$		& $6.12$	& $\begin{cases} 22.4 & t<9 \\ 3.31 & t\ge 9 \end{cases}$\\[.2ex]
Least squares error & $L_2$--diff & $8.2\cdot 10^3$	& $2.6\cdot 10^3$
\end{tabular}
\end{table}


\end{document}